\documentclass{llncs}
\usepackage{graphicx,amssymb,amsmath,cite}

\newcommand{\pair}[2]{$\langle #1, #2 \rangle$}
\newcommand{\pairb}[2]{\boldmath$\langle #1, #2 \rangle$}

\newcommand{\tp}{\hspace{12pt}}

\title{A Simple Linear-Space Data Structure for \\ Constant-Time 
Range Minimum Query\thanks{Work 
supported in part by the Natural Sciences and Engineering Research Council of 
Canada (NSERC).}}

\author{Stephane Durocher}

\institute{University of Manitoba, 
Winnipeg, Canada, \email{durocher@cs.umanitoba.ca}}

\begin{document}

\maketitle

\begin{abstract}
We revisit the range minimum query problem 
and present a new $O(n)$-space data structure that supports 
queries in $O(1)$ time.
Although previous data structures exist 
whose asymptotic bounds match ours,
our goal is to introduce a new solution that is
simple, intuitive, and practical
without increasing costs for query time or space.
\end{abstract}

\section{Introduction}
\label{sec:introduction}

\subsection{Motivation}
Along with the mean, median, and mode of a multiset, the minimum 
(equivalently, the maximum) is a fundamental 
statistic of data analysis for which efficient computation is necessary.
Given a list $A[0:n-1]$ of $n$ items drawn from a totally orderered set,
a {\em range minimum query (RMQ)} consists of an input pair of indices $(i, j)$ 
for which the minimum element of $A[i:j]$ must be returned.
The objective is to preprocess $A$ to construct a data structure
that supports efficient response to one or more subsequent range minimum
queries,
where the corresponding input parameters $(i,j)$ are provided at query time.

Although the complete set of possible queries can be precomputed and
stored using $\Theta(n^2)$ space, 
practical data structures require less storage 
while still enabling efficient response time.
For all $i$, if $i=j$, then a range query must report $A[i]$.
Consequently, any range query data structure 
for a list of $n$ items requires $\Omega(n)$ storage space in the 
worst case \cite{bose2005}.
This leads to a natural question: how quickly can an $O(n)$-space data
structure answer a range minimum query?

Previous $O(n)$-space data structures
exist that provide $O(1)$-time RMQ 
(e.g., \cite{bender2000,bender2005,berkman1993,fischer2006},
see Section~\ref{sec:relatedWork}).
These solutions typically require a transformation 
or invoke a property that enables the volume of stored precomputed data to be 
reduced while allowing constant-time access and RMQ computation.
Each such solution is a conceptual organization of the data into a compact 
table for efficient reference;
essentially, the algorithm reduces to a clever table lookup.
In this paper our objective is not to minimize the total number of bits
occupied by the data structure (our solution is not succinct) but
rather to present a simpler and more intuitive
method for organizing the precomputed data to support RMQ efficiently.
Our solution combines new ideas with 
techniques from various previous data structures:
van Emde Boas trees \cite{vanEmdeBoas1977},
resizable arrays \cite{munro1999},
range mode query \cite{krizanc2005},
one-sided RMQ \cite{bender2000},
and a linear-space data structure that supports RMQ in $O(\sqrt{n})$ time.
The resulting RMQ data structure matches previous optimal bounds of $O(n)$
space and $O(1)$ query time. 
Our data structure stores efficient representations of the data 
to permit direct lookup without requiring 
the indirect techniques employed by previous solutions, 
such as transformation to a lowest common ancestor query,
Cartesian trees, Eulerian tours, and the Four Russians speedup
(e.g., \cite{bender2000,bender2005,berkman1993,fischer2006}).

The RMQ problem is sometimes defined such that
a query returns only the index of the minimum element 
instead of the minimum element itself
(e.g., \cite{fischer2011}).
In this paper we require that the actual minimum element be returned.
As we discuss in Section~\ref{sec:relatedWork},
several succint data structures exist that support $O(1)$-time 
RMQ using only $O(n)$ bits of space. 
In order to return the minimum element in addition to its index,
any such data structure must also store the values from the input array $A$,
corresponding to a lower bound of $\Omega(n \log u)$ bits
or, equivalently, $\Omega(n)$ words of space in the worst case
(analogous lower bounds exist for other array range query problems,
e.g., see \cite{bose2005}).

\subsection{Definitions, Notation, and Model of Computation}
\label{sec:introduction.definitions}

We assume the RAM word model of computation with word size $\Theta(\log u)$,
where elements are drawn from a universe $U = \{ -u , \ldots , u-1 \}$
for some fixed $u \geq n$.
Unless stated otherwise, memory requirements are expressed in word-sized units.
We assume the usual set of $O(1)$-time primitive operations:
basic integer arithmetic 
(addition, subtraction, multiplication, division, and modulo),
bitwise logic, and bit shifts. 
We do not assume $O(1)$-time exponentiation nor, consequently, radicals.
When the base operand is a power of two and the result is an integer, however,
these operations can be computed using a bitwise left or right shift.
All arithmetic computations are on integers in $U$, and
integer division is assumed to return the floor of the quotient.
Finally, our data structure only requires finding the binary logarithm of
integers in the range $\{ 0 , \ldots , n\}$. 
Consequently, the complete set of values 
can be precomputed and stored in a table of size $O(n)$ 
to provide $O(1)$-time reference for the $\log$ and $\log\log$ operations 
at query time, regardless of whether logarithms are included in
the RAM model's primitive operations.

A common technique used in array range searching data structures 
(e.g., \cite{bender2000,krizanc2005})
is to partition the input array $A[0:n-1]$
into a sequence of $\lceil n/b \rceil$ blocks, each of size $b$ 
(except possibly for the last block whose size is $[(n-1) \bmod b] + 1$).
A query range $A[i:j]$ 
spans between $0$ and $\lceil n/b \rceil$ complete blocks.
We refer to the sequence of complete blocks contained within $A[i:j]$ 
as the {\em span}, 
to the elements of $A[i:j]$ that precede the span as the {\em prefix},
and to the elements of $A[i:j]$ that succeed the span as the {\em suffix}.
See Figure~\ref{fig:sqrtN}.
One or more of the prefix, span, and suffix may be empty.
When the span is empty, the prefix and suffix can lie either in
adjacent blocks, or in the same block;
in the latter case the prefix and suffix are equal.

We summarize the asymptotic resource requirements of a given 
RMQ data structure by the ordered pair \pair{f(n)}{g(n)},
where $f(n)$ denotes the storage space it requires
and $g(n)$ denotes its worst-case RMQ time.
Our discussion focuses primarily on these two measures of efficiency;
other measures of interest include the preprocessing time and the update time.
Note that similar notation is sometimes used 
to pair precomputation time and query time 
(e.g., \cite{bender2000,fischer2006}).

\section{Related Work}\label{sec:relatedWork}

Multiple \pair{\omega(n)}{O(1)} solutions are known, including 
precomputing RMQs for all query ranges in \pair{O(n^2)}{O(1)}, and
precomputing RMQs for all ranges of length $2^k$ for some $k\in \mathbb{Z}^+$
in \pair{O(n\log n)}{O(1)} 
(Sparse Table Algorithm) \cite{bender2000,fischer2006}.
In the latter case, a query is decomposed into two (possibly overlapping)
precomputed queries.
Similarly, \pair{O(n)}{\omega(1)} solutions exist,
including 
the \pair{O(n)}{O(\sqrt{n})} data structure
described in Section~\ref{sec:dataStructure.sqrtN}.

Several \pair{O(n)}{O(1)} RMQ data structures exist, 
many of which depend on the equivalence
between the range minimum query and lowest common ancestor (LCA) problems.
Harel and Tarjan \cite{tarjan1984} gave the first 
\pair{O(n)}{O(1)} solution to LCA.
Their solution was simplified by Schieber and Vishkin \cite{schieber1988}.
Berkman and Vishkin \cite{berkman1993} showed how to solve the LCA problem 
in \pair{O(n)}{O(1)} by transformation to RMQ using an Euler tour.
This method was simplified by Bender and Farach-Colton \cite{bender2000}
to give an ingenious solution which we briefly describe below.
Comprehensive overviews of previous 
solutions are given by Davoodi \cite{davoodi2011} and 
Fischer \cite{fischer2010}, respectively.

The array $A[0:n-1]$ can be transformed into a Cartesian tree
$\mathcal{C}(A)$ on $n$ nodes such that a RMQ on $A[i:j]$ 
corresponds to the LCA of the respective nodes associated with
$i$ and $j$ in $\mathcal{C}(A)$.
When each node in $\mathcal{C}(A)$ is labelled by its depth,
an Eulerian tour on $\mathcal{C}(A)$ 
(i.e., the depth-first traversal sequence on $\mathcal{C}(A)$)
gives an array 
$B[0:2n-2]$ for which any two adjacent values differ by $\pm 1$.
Thus, a LCA query on $\mathcal{C}(A)$ corresponds to a $\pm 1$-RMQ on $B$.
Array $B$ is partitioned into blocks of size $(\log n)/2$.
Separate data structures are constructed to answer queries that 
are contained within a single block of $B$ and those that
span multiple blocks, respectively.
In the former case, the $\pm 1$ property implies that
the number of unique blocks in $B$ is $O(\sqrt{n})$;
all $O(\sqrt{n}\log^2 n)$ RMQs on blocks of $B$ are precomputed
(the Four Russians technique).
In the latter case, 
a query can be decomposed into a prefix, span, and suffix
(see Section~\ref{sec:introduction.definitions}).
RMQs on the prefix and suffix are one-sided and can be found in $O(1)$ time
(see Section~\ref{sec:dataStructure.loglogN}).
The minimum of each block of $B$ is precomputed and stored in 
$A'[0:2n/\log n -1]$.
A RMQ on $A'$ (the minimum value in the span) can be found in \pair{O(n)}{O(1)}
using the \pair{O(n'\log n')}{O(1)} data structure mentioned above
due to the shorter length of $A'$ (i.e., $n' = 2n/\log n$).

Fischer and Heun \cite{fischer2006} use similar ideas to
give a \pair{O(n)}{O(1)}
solution to RMQ that applies the Four Russians technique to any array 
(i.e., it does not require the $\pm 1$ property)
on blocks of length $\Theta(\log n)$.
Yuan and Atallah \cite{yuan2010}
examine RMQ on multidimensional arrays and give
a new one-dimensional \pair{O(n)}{O(1)} solution
that uses a hierarchical binary decomposition 
of $A[0:n-1]$ into $\Theta(n)$ canonical intervals, 
each of length $2^k$ for some $k \in \mathbb{Z}^+$,
and precomputed queries within blocks
of length $\Theta(\log n)$ (similar to the Four Russians technique).

When only the index of the minimum is required, 
Sadakane \cite{sadakane2007} gives a succinct data structure
requiring $4n + o(n)$ bits that supports $O(1)$-time RMQ.
Fischer and Heun 
reduce the space requirements to $2n + o(n)$ \cite{fischer2007,fischer2011}.
Finally, the RMQ problem has been examined 
in the dynamic setting \cite{brodal2011b,davoodi2011},
in two and higher dimensions
\cite{brodal2011c,demaine2009,sadakane2007,fischer2007b,yuan2010},
and on trees and directed acyclic graphs
\cite{bender2005,brodal2011b,demaine2009}.

\section{A New \pairb{O(n)}{O(1)} RMQ Data Structure}
\label{sec:dataStructure}

The data structure is described in steps, starting with a previous
\pair{O(n)}{O(\sqrt{n})} data structure,
extending it to \pair{O(n\log\log n)}{O(\log\log n)} by applying the
technique recursively,
eliminating recursion to obtain \pair{O(n\log\log n)}{O(1)},
and finally reducing the space to \pair{O(n)}{O(1)}.
To simplify the presentation, suppose initially that the input array $A$
has size $n = 2^{2^k}$, for some $k \in \mathbb{Z}^+$; 
as described in Section~\ref{sec:dataStructure.generalize},
removing this constraint and generalizing to an arbitrary $n$ 
is easily achieved without any asymptotic increase in
time or space requirements.

\subsection{A \pairb{O(n)}{O(\sqrt{n})} RMQ Data Structure}
\label{sec:dataStructure.sqrtN}

The following \pair{O(n)}{O(\sqrt{n})} data structure 
is known in RMQ folklore (e.g., \cite{topcoder})
and has similar high-level structure to the $\pm 1$RMQ algorithm
of Bender and Farach-Colton \cite[Section 4]{bender2000}.
While subobtimal and often overlooked in favour of more efficient
solutions, this data structure forms the basis for our new \pair{O(n)}{O(1)}
data structure.

\begin{figure}
\centering
\includegraphics[width=0.75\linewidth]{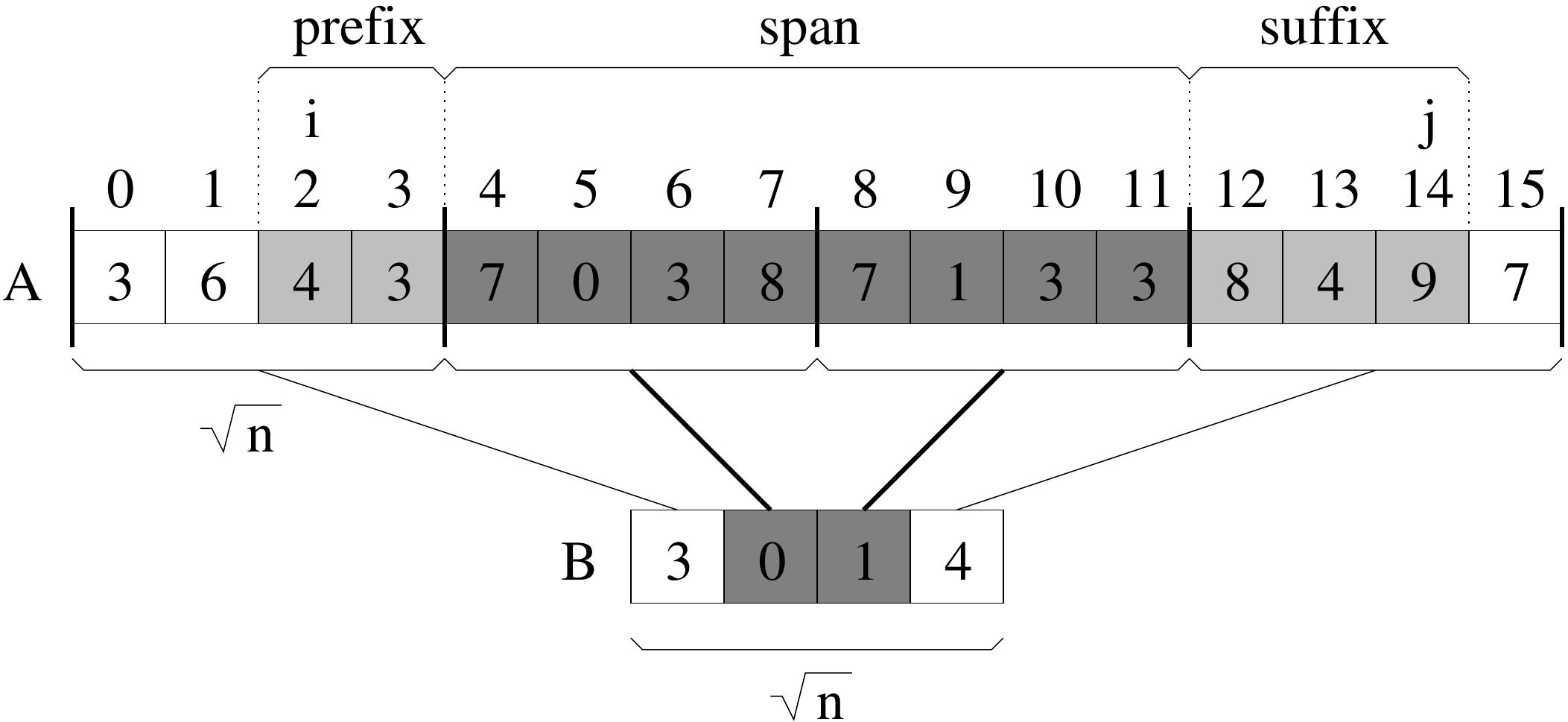}
\caption{A \pair{O(n)}{O(\sqrt{n})} data structure:
the array $A$ is partitioned into $\sqrt{n}$ blocks of size $\sqrt{n}$.
The range minimum of each block is precomputed and stored in array $B$.
A range minimum query $A[2:14]$ is processed by 
finding the minimum of the respective minima of 
the prefix $A[2:3]$, the span $A[4:11]$ (determined by examing array $B$), 
and the suffix $A[12:14]$.
In this example this corresponds to $\min\{3,0,4\} = 0$.}
\label{fig:sqrtN}
\end{figure}

The input array $A[0:n-1]$ is partitioned into $\sqrt{n}$ blocks of size 
$\sqrt{n}$.
The range minimum of each block is precomputed and stored in a table 
$B[0:\sqrt{n}-1]$.
See Figure~\ref{fig:sqrtN}.
A query range spans between zero and $\sqrt{n}$ complete blocks.
The minimum of the span is computed by iteratively examining the corresponding
values in $B$.
Similarly, the respective minima of the prefix and suffix are computed by 
iteratively examining their elements. 
The range minimum corresponds to the minimum of these three values.
Since the prefix, suffix, and array $B$ each contain
at most $\sqrt{n}$ elements, 
the worst-case query time is $\Theta(\sqrt{n})$.
The total space required by the data structure is $\Theta(n)$.
Precomputation requires only a single pass over the input array 
in $\Theta(n)$ time.
Updates require $\Theta(\sqrt{n})$ time in the worst case;
whenever an array element equal to its block's minimum is increased, 
the block must be scanned to identify the new minimum.

\subsection{A \pairb{O(n\log\log n)}{O(\log\log n)} RMQ Data Structure}
\label{sec:dataStructure.loglogN}

One-sided range minimum queries 
are trivially precomputed\cite{bender2000} 
and stored in arrays $C$ and $C'$, each of size $n$, where for each $i$,
\begin{align}
C[i] & = \begin{cases} \min \{ A[i], C[i-1]\} & \text{if~} i > 0, \\
A[0] & \text{if~} i = 0 , \end{cases} 
\nonumber \\
\text{and} \quad
C'[i] & = \begin{cases} \min \{ A[i], C'[i+1]\} & \text{if~} 
i < n-1, \\
A[n-1] & \text{if~} i = n-1 . \end{cases}
\label{eqn:oneSided}
\end{align}
Any subsequent one-sided range minimum query on $A[0:j]$ or $A[j:n-1]$
can be answered in $O(1)$ time by referring to $C[j]$ or $C'[j]$.

The \pair{O(n)}{O(\sqrt{n})} solution discussed 
in Section~\ref{sec:dataStructure.sqrtN}
includes three range minimum queries on subproblems of size $\sqrt{n}$,
of which at most one is two-sided.
In particular, if the span is non-empty, 
then the query on array $B$ is two-sided, 
and the queries on the prefix and suffix are one-sided.
Similarly, if the query range is contained in a single block, 
then there is a single two-sided query and no one-sided queries.
Finally, if the query range intersects exactly two blocks,
then there are two one-sided queries (one each for the prefix and suffix)
and no two-sided queries.

Thus, upon adding arrays $C$ and $C'$ to the data structure,
at most one of the three (or fewer) subproblems requires 
$\omega(1)$ time to identify its range minimum.
This search technique can be applied recursively on two-sided queries.
By limiting the number of recursive calls to at most one 
and by reducing the problem size by an exponential factor of $1/2$
at each step of the recursion,
the resulting query time is bounded by the following recurrence
(similar to that achieved by van Emde Boas trees 
\cite{vanEmdeBoas1977}):
\begin{align}
T(n) & \leq 
\begin{cases} T(\sqrt{n}) + O(1) & \text{if~} n > 2, \\
O(1) & \text{if~} n \leq 2 \end{cases}
\nonumber \\
& \in O(\log \log n) . 
\label{eqn:loglogn}
\end{align}

Each step invokes at most one recursive range minimum query on a subarray of 
size $\sqrt{n}$. 
Each recursive call is one of two types: 
i) a recursive call on array $B$ 
(a two-sided query to compute the range minimum of the span) 
or ii) a recursive call on the entire query range 
(contained within a single block).

Recursion can be avoided entirely for determining the minimum of the span
(a recursive call of the first type).
Since there are $\sqrt{n}$ blocks, ${\sqrt{n} + 1 \choose 2} < n$ 
distinct spans are possible.
As is done in the range mode query data structure of 
Krizanc et al.~\cite{krizanc2005},
the minimum of each span can be precomputed and stored in a table $D$ of size 
$n$.
Any subsequent range minimum query on a span can be answered in $O(1)$ time
by reference to table $D$.
Consequently, tables $C$ and $D$ suffice, and table $B$ can be eliminated.

The result is a hierarchical data structure containing 
$\log\log n + 1$ levels\footnote{Throughout this manuscript,
$\log a$ denotes the binary logarithm $\log_2 a$.} 
which we number $0, \ldots, \log\log n$, 
where the $x$th level\footnote{Level $\log\log n$ is included 
for completeness since we refer to the size of the parent of blocks on level 
$x$, for each $x \in \{0, \ldots , \log\log n-1\}$.
The only query that refers to level $\log\log n$ directly is the complete array:
$i = 0$ and $j=n-1$.
The minimum for this singular case can be stored using $O(1)$ space and updated
in $O(\sqrt{n})$ time as described in Section~\ref{sec:dataStructure.sqrtN}.}
is a sequence of $b_x(n) = n \cdot 2^{-2^x}$ blocks 
of size $s_x(n) = n / b_x(n) = 2^{2^x}$.
See Table~\ref{tab:levels}.

\begin{table}
\centering
\begin{tabular}%
{c@{~~~}c@{~~~}c@{~~~}c@{~~~}c@{~~~}c@{~~~}c@{~~~}c@{~~~}c@{~~~}c}
level $x$ & 0 & 1 & 2 & $\ldots$ & $i$ & $\ldots$ & $\log\log n - 2$ 
& $\log\log n - 1$ & $\log\log n$ \\
$b_x(n)$ & $n/2$ & $n/4$ & $n/16$ & $\ldots$ & $n 2^{-2^i}$ & $\ldots$ 
& $n^{3/4}$ & $\sqrt{n}$ & 1 \\
$s_x(n)$ & 2 & 4 & 16 & $\ldots$ & $2^{2^i}$ & $\ldots$ & $n^{1/4}$ & $\sqrt{n}$
& $n$ \\
\\
\end{tabular}
\caption{The $x$th level is a sequence of $b_x(n)$ blocks of size $s_x(n)$.}
\label{tab:levels}
\end{table}

Generalizing \eqref{eqn:oneSided}, the new arrays $C_x$ and $C'_x$ 
are defined by
\begin{align}
C_x[i] & = \begin{cases} \min \{ A[i], C_x[i-1]\} 
& \text{if~} i \neq 0 \bmod s_x(n), \\
A[i] & \text{if~} i = 0 \bmod s_x(n), \\ \end{cases} 
\nonumber \\
\text{and} \quad
C_x'[i] & = \begin{cases} \min \{ A[i], C_x'[i+1]\} 
& \text{if~} (i+1) \neq 0 \bmod s_x(n), \\
A[i] & \text{if~} (i+1) = 0 \bmod s_x(n). \\ \end{cases} 
\nonumber
\end{align}

We refer to a sequence of blocks on level $x$ that are contained in a
common block on level $x+1$ as {\em siblings} and to the common block
as their {\em parent}.
Each block on level $x+1$ is a parent to $s_{x+1}(n) / s_x(n) = s_x(n)$ 
siblings on level $x$.
Thus, any query range contained in some block at level $x+1$ 
covers at most $s_x(n)$ siblings at level $x$,
resulting in $\Theta(s_x(n)^2) = \Theta(s_{x+1}(n))$ 
distinct possible spans within a block at level $x+1$
and $\Theta(s_{x+1}(n) \cdot b_{x+1}(n)) = \Theta(n)$ 
total distinct possible spans at level $x+1$, 
for any $x \in \{0, \ldots, \log\log n - 1\}$.
These precomputed range minima are stored in table $D$, 
such that for every $x \in \{0, \ldots, \log\log n - 1\}$, 
every $b \in \{ 0, \ldots, b_{x+1}(n) - 1\}$,
and every $\{i,j\} \subseteq \{ 0 , \ldots, s_x(n) - 1\}$,
$D_x[b][i][j]$ stores the minimum of the span 
$A[b \cdot s_{x+1}(n) + i \cdot s_x(n): b \cdot s_{x+1}(n) + (j+1) s_x(n) - 1]$.

This gives the following recursive algorithm
whose worst-case time is bounded by \eqref{eqn:loglogn}:

\bigskip

\noindent
\begin{tabular}{@{}r@{\hspace{3pt}}l@{\hspace{3pt}}l}
\multicolumn{3}{l}{\bf Algorithm 1}\\
& ~ \\
\multicolumn{3}{l}{\sc RMQ$(i, j)$} \\
1 & {\bf if} $i = 0$ {\bf and} $j=n-1$ 
& // query is entire array \\
2 & \tp {\bf return} $minA$ 
& // precomputed array minimum \\
3 & {\bf else} \\
4 & \tp {\bf return} $RMQ(\log\log n-1, i, j)$ 
& // start recursion at top level \\
& ~ \\
\multicolumn{3}{l}{\sc RMQ$(x, i, j)$} \\
1 & {\bf if} $x > 0$ \\
2 & \tp $b_i \leftarrow \lfloor i / s_x(n) \rfloor$ &
// blocks containing $i$ and $j$ \\
3 & \tp $b_j \leftarrow \lfloor j / s_x(n) \rfloor$ \\
4 & \tp {\bf if} $b_i = b_j$ &
// $i$ and $j$ in same block at level $x$ \\
5 & \tp\tp {\bf return} RMQ$(x-1,i,j)$ &
// two-sided recursive RMQ: $T(\sqrt{n})$ time \\
6 & \tp {\bf else if} $b_j - b_i \geq 2$ &
// span is non-empty \\
7 & \tp\tp $b \leftarrow i \bmod s_{x+1}(n)$ \\
8 & \multicolumn{2}{l}{\tp\tp {\bf return} 
$\min \{ C_x'[i], C_x[j], D_x[b][b_i+1][b_j-1] \}$} \\
& \multicolumn{2}{l}{\tp\tp // 2 one-sided RMQs + precomputed span: 
$O(1)$ time} \\ 
9 & \tp {\bf else} \\
10 & \tp\tp {\bf return} $\min \{ C_x'[i], C_x[j] \}$ &
// 2 one-sided RMQs: $O(1)$ time \\ 
11 & {\bf else} \\
12 & \tp {\bf return} $\min\{ A[i], A[j] \}$ &
// base case (block size $\leq 2$): $O(1)$ time \\
\end{tabular}

\bigskip

The space required by array $D_x$ for each level $x < \log\log n$ is
\[ O\left( s_x(n)^2 \cdot b_{x+1}(n) \right)
= O\left( s_{x+1}(n) \cdot b_{x+1}(n) \right) = O(n) . \]
Since arrays $C_x$ and $C_x'$ also require $O(n)$ space at each level, 
the total space required is $O(n)$ per level,
resulting in $O(n \log\log n)$ total space for the complete data structure.

For each level $x < \log\log n$, precomputing arrays $C_x$, $C_x'$, and $D_x$ 
is easily achieved in $O(n \cdot s_x(n)) = O(n \cdot 2^{2^x})$ time 
per level, or $O(n^{3/2})$ total time. 
Each update requires $O(s_x(n))$ time per level, or $O(\sqrt{n})$ total time
per update.

\subsection{A \pairb{O(n\log\log n)}{O(1)} RMQ Data Structure}
\label{sec:dataStructure.constant1}

Each step of Algorithm 1 described in Section~\ref{sec:dataStructure.loglogN}
invokes at most one recursive call on a subarray whose size
decreases exponentially at each step.
Specifically, the only case requiring $\omega(1)$ time
occurs when the query range is contained within a single block of the current
level.
In this case, no actual computation or table lookup occurs locally;
instead, the result of the recursive call is returned directly
(see Line 5 of Algorithm 1).
As such, the recursion can be eliminated by jumping directly to the
corresponding level of the data structure at which the recursion terminates,
that is, the highest level of the data structure
for which the query range is not contained in a single block.
Any such query can be answered in $O(1)$ time using a combination of
at most three references to arrays $C$ and $D$
(see Lines 8 and 10 of Algorithm 1).
We refer to the corresponding level of the data structure as the
{\em query level}, whose index we denote by $\ell$.

More precisely,
Algorithm 1 makes a recursive call 
whenever $b_i = b_j$, where $b_i$ and $b_j$ denote the respective indices of 
the blocks containing $i$ and $j$ in the current level
(see Line 5 of Algorithm 1).
Thus, we seek to identify the highest level for which $b_i \neq b_j$.
In fact, it suffices to identify the highest level 
$\ell \in \{0, \ldots, \log\log n-1\}$ for which
no query of size $j-i+1$ can be contained within a single block.
While the query could span the boundary of (at most) two adjacent blocks 
at higher levels,
it must span at least two blocks at all levels less than or equal to $\ell$.
In other words, the size of the query range is bounded by 
\begin{alignat*}{3}
&& s_\ell(n) & < & j - i + 1 & \leq s_{\ell+1}(n) 
\\
&\Leftrightarrow~ & 2^{2^\ell} & < & j - i + 1 & \leq 2^{2^{\ell+1}} 
\\
& \Leftrightarrow~ &
\log\log(j-i+1) - 1 & \leq & \ell & < \log\log(j-i+1) 
\\
& \Rightarrow~ & &&
\ell & = \lfloor \log\log(j-i) \rfloor .
\nonumber 
\end{alignat*}

As discussed in Section~\ref{sec:introduction.definitions},
since we only require finding binary logarithms of positive integers
up to $n$, these values can be precomputed and stored in a table of size $O(n)$.
Consequently, the value $\ell$ can be computed in $O(1)$ time
at query time, where each logarithm is found by a table lookup. 

This gives the following simple algorithm
whose worst-case running time is constant
(note the absence of loops or recursive calls):

\bigskip

\noindent
\begin{tabular}{@{}r@{\hspace{3pt}}l@{\hspace{3pt}}l}
\multicolumn{3}{l}{\bf Algorithm 2}\\
& ~ \\
\multicolumn{3}{l}{\sc RMQ$(i, j)$} \\
1 & {\bf if} $i = 0$ {\bf and} $j=n-1$ 
& // query is entire array \\
2 & \tp {\bf return} $minA$ 
& // precomputed array minimum \\
3 & {\bf else if $j - i \geq 2$} \\
4 & \tp $\ell \leftarrow \lfloor \log\log (j-i)\rfloor$ \\
5 & \tp $b_i \leftarrow \lfloor i / s_\ell(n) \rfloor$ \\
6 & \tp $b_j \leftarrow \lfloor j / s_\ell(n) \rfloor$ &
// blocks containing $i$ and $j$ \\
7 & \tp {\bf if} $b_j - b_i \geq 2$ & // span is non-empty \\
8 & \tp\tp $b \leftarrow i \bmod s_{\ell+1}(n)$ \\
9 & \multicolumn{2}{l}{\tp\tp {\bf return} 
$\min \{ C_\ell'[i], C_\ell[j], D_\ell[b][b_i+1][b_j-1] \}$} \\
& \multicolumn{2}{l}{\tp\tp
// 2 one-sided RMQs + precomputed span: $O(1)$ time} \\ 
10 & \tp {\bf else} \\
11 & \tp\tp {\bf return} $\min \{ C_\ell'[i], C_\ell[j] \}$ &
// 2 one-sided RMQs: $O(1)$ time \\ 
12 & {\bf else} \\
13 & \tp {\bf return} $\min \{A[i], A[j]\}$ &
// query contains $\leq 2$ elements \\
\end{tabular}

\bigskip

Although the query algorithm differs from Algorithm 1,
the data structure remains unchanged except for the addition of
precomputed values for logarithms
which require $O(n)$ additional space total space.
As such, the space remains $O(n \log\log n)$ while the query time
is reduced to $O(1)$ in the worst case.
Precomputation and update times remain $O(n^{3/2})$ and $O(\sqrt{n})$,
respectively.

\subsection{A \pairb{O(n)}{O(1)} RMQ Data Structure}
\label{sec:dataStructure.constant2}

The data structures described in Sections~\ref{sec:dataStructure.loglogN}
and~\ref{sec:dataStructure.constant1}
store exact precomputed values in arrays $C_x$, $C'_x$, and $D_x$.
That is, for each $a$ and each $x$, $C_x[a]$ stores $A[b]$ for some $b$
(similarly for $C_x'$ and $D_x$).
If the array $A$ is accessible during a query, 
then it suffices to store the relative index $b-a$ instead of storing $A[b]$.
Thus, $C_x[a]$ stores $b-a$ and the returned value is 
$A[C_x[a] + a] = A[(b-a)+a] = A[b]$.
Since the range minimum is contained in the query range $A[i:j]$
we get that $\{a, b\} \subseteq \{i, \ldots, j\}$ and, therefore,
\[ |b-a| \leq j-i+1 \leq s_{\ell+1}(n) . \]
Consequently, for each level $x$, $\log(s_{x+1}(n)) = 2^{x+1}$ bits 
suffice to encode any value stored in $C_x$, $C_x'$, or $D_x$.
Therefore, for each level $x$, each table $C_x$, $C_x'$, and $D_x$
can be stored using $O(n \cdot 2^{x+1})$ bits.
Observe that
\begin{equation}
\sum_{x=0}^{\log\log n-1} 2^{x+1} < 2\log n 
\text{~~~and, similarly, ~~}
\sum_{x=0}^{\log\log n-1} n \cdot 2^{x+1} < 2n\log n . 
\label{eqn:compact}
\end{equation}
Consequently, the total space occupied by the tables $C_x$, $C_x'$, and $D_x$
can be compacted into $O(n\log n)$ bits or, equivalently, $O(n)$ words of space.
We now describe how to store this compact representation
to enable efficient access.
For each $i \in \{0, \ldots , n - 1\}$, 
the values $C_0[i], \ldots , C_{\log\log n-1}[i]$ can be stored 
in two words by \eqref{eqn:compact}.
Specifically, the first word stores $C_{\log\log n-1}[i]$
and for each $x \in \{0, \ldots , \log\log n - 2\}$, 
bits $2^{x+1}-1$ through $2^{x+2}-2$ store the value $C_x[i]$.
Thus, all values $C_0[i], \ldots , C_{\log\log n - 2}[i]$ are stored using
\[ \sum_{i=0}^{\log\log n - 2} 2^{x+1} = \log n - 2 < \log u \]
bits, i.e., a single word, 
where $\log u$ denotes the word size under the RAM model.
The value $C_x[i]$ can be retrieved using a bitwise left shift followed by 
a right shift or, alternatively, 
a bitwise logical AND with the corresponding sequence of consecutive 1 bits 
(all $O(\log\log n)$ bit sequences can be precomputed).
An analogous argument applies to the arrays $C_x'$ and $D$, resulting
in $O(n)$ space for the complete data structure.

To summarize, the query algorithm is unchanged from Algorithm 2
and the corresponding query time remains constant, 
but the data structure's required space is reduced to $O(n)$.
Precomputation and update times remain $O(n^{3/2})$ and $O(\sqrt{n})$,
respectively. This gives the following lemma:

\begin{lemma}
\label{lem:constant2}
Given any $n = 2^{2^k}$ for some $k\in \mathbb{Z}^+$
and any array $A[0:n-1]$, Algorithm 2 supports range minimum queries on 
$A$ in $O(1)$ time using a data structure of size $O(n)$.
\end{lemma}

\subsection{Generalizing to an Arbitrary Array Size \boldmath$n$}
\label{sec:dataStructure.generalize}

To simplify the presentation in 
Sections~\ref{sec:dataStructure.sqrtN} to~\ref{sec:dataStructure.constant2}
we assumed that the input array had size $n=2^{2^k}$ 
for some $k \in \mathbb{Z}^+$. As we show in this section,
generalizing the data structure to an arbitrary positive integer $n$ 
while maintaining the same bounds on space and time is straightforward.

Let $m$ denote the largest value no larger than $n$ for which 
Lemma~\ref{lem:constant2} applies. That is,
\begin{align}
&& m & = 2^{2^{\lfloor \log \log n \rfloor}} 
\nonumber \\
\Rightarrow && m & \leq n < m^2
\nonumber \\
\Rightarrow && n / m & < \sqrt{n} .
\label{eqn:defM}
\end{align}
Define a new array $A'[0:n'-1]$, where $n' = m \lceil n/m \rceil$, 
that corresponds to the array $A$
padded with dummy data\footnote{For implementation, it suffices to store $u-1$
(the largest value in the universe $U$)
instead of $+\infty$ as the additional values.}
to round up to the next multiple of $m$. Thus,
\[ \forall i \in \{0, \ldots, n'-1\} , \
A'[i] = \begin{cases} A[i] & \text{if~} i < n \\
+\infty & \text{if~} i \geq n . \end{cases} \]
Since $n' = 0 \bmod m$,
partition array $A'$ into a sequence of blocks of size $m$. 
The number of blocks in $A'$ is $\lceil n / m \rceil < \lceil \sqrt{n} \rceil$.

By \eqref{eqn:defM} and Lemma~\ref{lem:constant2},
for each block we can construct a data structure 
to support range mode query on that block in $O(1)$ time using $O(m)$ space
per block.
Therefore, the total space required by all blocks in $A'$ is
$O( \lceil n/m \rceil \cdot m) = O(n)$.
Construct arrays $C$, $C'$, and $D$ as before on the top level of array $A'$
using the blocks of size $m$.
The arrays $C$ and $C'$ each require $O(n') = O(n)$ space.
The array $D$ requires $O(\lceil n/m \rceil^2) \subseteq O(n)$ space
by \eqref{eqn:defM}.
Therefore, the total space required by 
the complete data structure remains $O(n)$.

Each query is performed as in Algorithm 2, except that references to $C$, $C'$,
and $D$ at the top level access the corresponding arrays (which are stored 
separately from $C_x$, $C_x'$, and $D_x$ for the lower levels).
Therefore, the query time is increased by a constant factor for the 
first step at the top level, and the total query time remains $O(1)$.

This gives the following theorem:

\begin{theorem}[Main Result]
\label{thm:main}
Given any $n \in \mathbb{Z}^+$,
and any array $A[0:n-1]$, Algorithm 2 supports range minimum queries on 
$A$ in $O(1)$ time using a data structure of size $O(n)$.
\end{theorem}

\section{Discussion and Directions for Future Work}

{\bf Succinctness.}
The data structure presented in this paper uses $O(n)$ words of space.
It is not currently known whether its space can be reduced to $O(n)$ bits 
if a RMQ returns only the index of the minimum element.
As suggested by Patrick Nicholson (personal communication, 2011),
each array $C_x$ and $C_x'$ can be stored
using binary rank and select data structures in $O(n)$ bits of space 
(e.g., \cite{munro1996}).
That is, we can support references to $C_x$ and $C_x'$ in constant time
using $O(n)$ bits of space per level or $O(n \log\log n)$ total bits.
It is not known whether the remaining components of the data structure
can be compressed similarly, or whether the space can be reduced further
to $O(n)$ bits.

\medskip

\noindent
{\bf Higher Dimensions.}
As shown by Demaine et al.\ \cite{demaine2009}, RMQ data structures based
on Cartesian trees cannot be generalized to two or higher dimensions.
The data structure presented in this paper does not involve Cartesian trees.
Although it is possible that some other constraint may preclude generalization
to higher dimensions, this remains to be examined.

\medskip

\noindent
{\bf Dynamic Data.}
As described, our data structure structure requires $O(\sqrt{n})$ time 
per update in the worst case.
It is not known whether the data structure can be modified 
to support efficient queries and updates without increasing space.

\section*{Acknowledgements}

The author thanks Timothy Chan and Patrick Nicholson 
with whom this paper's results were discussed.
The author also thanks the students of his senior undergraduate 
class in advanced data structures at the University of Manitoba
with whom discussion of range searching in arrays
inspired him to re-examine the range minimum query problem.

\bibliographystyle{plain}
\bibliography{rmqAbbrev}

\begin{thebibliography}{10}

\bibitem{fischer2007b}
A.~Amir, J.~Fischer, and M.~Lewenstein.
\newblock Two-dimensional range minimum queries.
\newblock In {\em Proc.\ CPM}, volume 4580 of {\em LNCS}, pages 286--294.
  Springer, 2007.

\bibitem{bender2000}
M.~A. Bender and M.~Farach-Colton.
\newblock The {LCA} problem revisited.
\newblock In {\em Proc.\ LATIN}, volume 1776 of {\em LNCS}, pages 88--94.
  Springer, 2000.

\bibitem{bender2005}
M.~A. Bender, M.~Farach-Colton, G.~Pemmasani, S.~Skiena, and P.~Sumazin.
\newblock Lowest common ancestors in trees and directed acyclic graphs.
\newblock {\em J.\ Alg.}, 57(2):75--94, 2005.

\bibitem{berkman1993}
O.~Berkman and U.~Vishkin.
\newblock Recursive star-tree parallel data structures.
\newblock {\em {SIAM} J.\ Comp.}, 22(2):221--242, 1993.

\bibitem{bose2005}
P.~Bose, E.~Kranakis, P.~Morin, and Y.~Tang.
\newblock Approximate range mode and range median queries.
\newblock In {\em Proc.\ STACS}, volume 3404 of {\em LNCS}, pages 377--388.
  Springer, 2005.

\bibitem{brodal2011c}
G.~S. Brodal, P.~Davoodi, and S.~S. Rao.
\newblock On space efficient two dimensional range minimum data structures.
\newblock {\em Algorithmica}, 2011.
\newblock To appear.

\bibitem{brodal2011b}
G.~S. Brodal, P.~Davoodi, and S.~S. Rao.
\newblock Path minima queries in dynamic weighted trees.
\newblock In {\em Proc.\ WADS}, volume 6844 of {\em LNCS}, pages 290--301.
  Springer, 2011.

\bibitem{munro1999}
A.~Brodnik, S.~Carlsson, E.~D. Demaine, J.~I. Munro, and R.~Sedgewick.
\newblock Resizable arrays in optimal time and space.
\newblock In {\em Proc.\ WADS}, volume 1663 of {\em LNCS}, pages 27--48.
  Springer, 1999.

\bibitem{davoodi2011}
P.~Davoodi.
\newblock {\em Data Structures: Range Queries and Space Efficiency}.
\newblock PhD thesis, Aarhus University, 2011.

\bibitem{demaine2009}
E.~Demaine, G.~M. Landau, and O.~Weimann.
\newblock On {Cartesian} trees and range minimum queries.
\newblock In {\em Proc.\ ICALP}, volume 5555 of {\em LNCS}, pages 341--353.
  Springer, 2009.

\bibitem{vanEmdeBoas1977}
P.~van Emde~Boas.
\newblock Preserving order in a forest in less than logarithmic time and linear
  space.
\newblock {\em Inf.\ Proc.\ Let.}, 6(3):80--82, 1977.

\bibitem{fischer2010}
J.~Fischer.
\newblock Optimal succinctness for range minimum queries.
\newblock In {\em Proc.\ LATIN}, volume 6034 of {\em LNCS}, pages 158--169.
  Springer, 2010.

\bibitem{fischer2006}
J.~Fischer and V.~Heun.
\newblock Theoretical and practical improvements on the {RMQ}-problem with
  applications to {LCA} and {LCE}.
\newblock In {\em Proc.\ CPM}, volume 4009 of {\em LNCS}, pages 36--48.
  Springer, 2006.

\bibitem{fischer2007}
J.~Fischer and V.~Heun.
\newblock A new succinct representation of {RMQ}-information and improvements
  in the enhanced suffix array.
\newblock In {\em Proc.\ ESCAPE}, volume 4614 of {\em LNCS}, pages 459--470.
  Springer, 2007.

\bibitem{fischer2011}
J.~Fischer and V.~Heun.
\newblock Space-efficient preprocessing schemes for range minimum queries on
  static arrays.
\newblock {\em {SIAM} J.\ Comp.}, 40(2):465--492, 2011.

\bibitem{tarjan1984}
D.~Harel and R.~E. Tarjan.
\newblock Fast algorithms for finding nearest common ancestors.
\newblock {\em {SIAM} J.\ Comp.}, 13(2):338--355, 1984.

\bibitem{krizanc2005}
D.~Krizanc, P.~Morin, and M.~Smid.
\newblock Range mode and range median queries on lists and trees.
\newblock {\em Nordic J.\ Comp.}, 12:1--17, 2005.

\bibitem{munro1996}
J.~I. Munro.
\newblock Tables.
\newblock In V.~Chandru and V.~Vinay, editors, {\em Foundations of Software
  Technology and Theoretical Computer Science}, volume 1180 of {\em LNCS},
  pages 37--42. Springer, 1996.

\bibitem{topcoder}
D.~Pasail\u{a}.
\newblock Range minimum query and lowest common ancestor.
\newblock
  http://www.topcoder.com/tc?module=Static\&d1=tutorials\&d2=lowestCommonAnces%
tor.

\bibitem{sadakane2007}
K.~Sadakane.
\newblock Succinct data structures for flexible text retrieval systems.
\newblock {\em J.\ Disc.\ Alg.}, 5:12--22, 2007.

\bibitem{schieber1988}
B.~Schieber and U.~Vishkin.
\newblock On finding lowest common ancestors: Simplification and
  parallelization.
\newblock {\em {SIAM} J.\ Comp.}, 17(6):1253--1262, 1988.

\bibitem{yuan2010}
H.~Yuan and M.~J. Atallah.
\newblock Data structures for range minimum queries.
\newblock In {\em Proc.\ SODA}, pages 150--160, 2010.

\end{thebibliography}

\end{document}